\documentstyle[11pt,paspconf,epsf]{article}

\begin{document}

\title{The Metallicity Effect on the Cepheid P-L relation from SMC Cepheids}

\author{Jesper Storm}
\affil{Astrophysikalisches Institut Potsdam, Germany}
\author{Bruce W. Carney}
\affil{University of North Carolina at Chapel Hill, North Carolina, U.S.A.}
\author{Anne M. Fry}
\affil{Case Western Reserve University, Ohio, U.S.A.}

\begin{abstract}
  We present results based on a Baade-Wesselink analysis of Cepheids in
the Small Magellanic Cloud. The Baade-Wesselink analysis provides individual
luminosities for these metal-poor Cepheids which combined with
recent Baade-Wesselink results from Gieren et al. (1998) on solar metallicity
Galactic Cepheids constrain the metallicity effect on the zero-point
of the Cepheid P-L relation.
A preliminary analysis leads to an effect of 
$\Delta \mbox{$M_{V}\,$} / \Delta {\mbox{[Fe/H]}\,} =
-0.45 \pm 0.15$, metal-rich Cepheids being brighter, in good agreement
with several recent independent determinations. An effect of this
magnitude reduces significantly the current disagreement between the
long and the short distance estimates to the LMC, and favors a shorter
value.
\end{abstract}


\section{Introduction}

  With the large efforts currently being carried out with the Hubble
Space Telescope to measure Cepheids in distant galaxies to define
the extra-galactic distance scale, it is becoming more important than
ever to proper understand the Cepheid P-L relation. Not only is it
important to determine the zero-point of the relation but in particular it
is important to determine the effect of metallicity on this zero
point as the LMC, which is used as a reference, is more metal-poor than
most galaxies available for
study with HST. The Baade-Wesselink method provides luminosities for individual
stars and by analyzing stars spanning a range of metallicities
the associated luminosity shift can be determined.

\section{The Data}
  Five SMC Cepheids with periods between 13.5 and 16.7 days spanning a
range in $(B-V)$ color from 0.52 to 0.80 have been observed. The stars
are located in the south-western part of the SMC, about 1 degree from
the main body of the galaxy.

  The dataset consists of light curves ($\approx 40$ points per star)
in BVRI based on CCD photometry from Las Campanas, CTIO and ESO as
well as radial velocity curves ($\approx 25$ points per star)
based on echelle spectroscopy from Las Campanas and supplemented at ESO.

  One star, HV1328, exhibits a very low amplitude in the radial velocity
curve and we have so far not been able to successfully
derive a distance to this star. This star has been disregarded in
the following leaving us with a sample of four stars.
  
\section{The Method}

  We have used the recent calibration by Fouqu\'e and Gieren (1997) of
the Barnes-Evans (1976) realization of the Baade-Wesselink method. This
calibration is based on the most recent interferometric diameters of
giants and super-giants. They calibrate the surface brightness parameter
$F_V$ as a function of the color-index $(V-R_J)$ and find the relation:

\begin{equation}
F_V = 3.947 - 0.380(V-R_J)_0
\end{equation}

  The surface brightness parameter $F_V$ together with the visual
magnitude gives the angular diameter $\theta$ of the star through the
relation:

\begin{equation}
F_V = 4.2207 - 0.1V - 0.5 \log \theta
\end{equation}

  From the integrated radial velocity curve, the linear displacement of
the stellar surface can be determined as a function of phase and the
method determines the stellar radius and distance to the star
from the fit to the pairs of linear displacement and angular
diameters at different phases (see panel $b$ in Figure~\ref{fig-svb}). 

  Gieren et al. (1997) show that the calibration for the $(V-K)$ index
is in excellent
agreement with the main-sequence fitting results to the galactic
calibrating Cepheids in open clusters and associations.

\section{Metallicity effect on the Barnes-Evans method}

  The surface brightness parameter is basically a temperature measure and
it is based on a color-index. Consequently it is very likely that there
is a certain
dependence on abundance on the relation. As the SMC Cepheids are of
significantly lower metallicity than the
Cepheids in the Galaxy and the calibrating stars, we have to determine
the size of this effect and correct the calibration accordingly 
before we apply it to the SMC stars. 

  We have used Kurucz model atmospheres at fixed gravity to modify the
calibration above for different assumed metallicities. These modified
calibrations have then been used to determine Baade-Wesselink
luminosities for our sample of stars.

The trend is largely linear and shows an effect of the order
$-0.2$~mag per dex but with some differences from star to star.

\section{Absolute physical parameters}

\begin{figure}
\plotfiddle{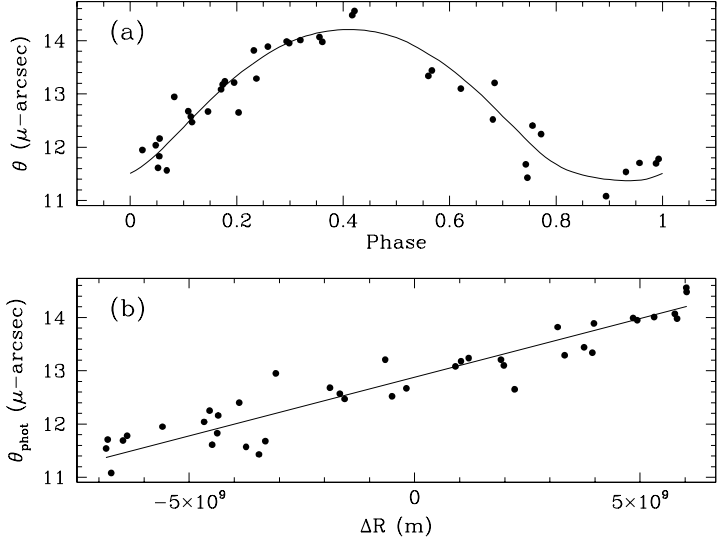}{6cm}{0}{50}{50}{-150}{0}
\caption{Panel $a$ shows the angular diameter variation in micro arc-seconds
as a function of phase for the star HV1333. Panel $b$ shows the fit to 
the data pairs of angular
diameter variation versus absolute diameter variation.}
\label{fig-svb}
\end{figure}

  Following Luck et al. (1998) we adopt an abundance difference of $-0.7$
between the SMC and the Galaxy. Using the corrected calibration of the
surface brightness parameter we can apply the Baade-Wesselink method to
our stars to determine individual radii, distances and absolute
magnitudes.

One critical point in the implementation of the 
Barnes-Evans method using the $(V-R_J)$ index is the choice of fitting
method as there is a significant scatter in the data points. However, as
we here are interested in the comparison with the absolute magnitudes of
the Galactic Cepheids determined by Gieren et al. (1998) using the very same
calibration, we have to adopt the same fitting scheme as they have used,
i.e. performing linear regression with the angular diameter as the
dependent variable.  In this way we derive the values in
Table \ref{tab.FGno_bisec}. The typical uncertainty on the distance modulus
to a single star is about $0.12$~mag. An example of the fit to the angular
diameter as a function of linear radius displacement is shown in
panel $b$ of Figure~\ref{fig-svb}.

Unfortunately Gieren et al. (1997) find a
systematic offset of 16\% or 0.32~mag between the absolute magnitudes
determined on the basis of the $(V - R_J)$ and the $(V - K)$ color
indices. As they also find that the $(V-K)$ results are superior in accuracy
and precision over the $(V-R_J)$ results, we have to
correct for this offset before comparing with the results from Gieren
et al. (1998). Work is in progress together with Gieren and
collaborators to obtain
near-IR data for the SMC Cepheids so we can repeat the analysis without
the added uncertainty from this offset.

\begin{table}
\caption{\label{tab.FGno_bisec}Physical parameters (uncorrected, see
text) resulting from the
Baade-Wesselink analysis of the stars. The angular diameter was used as
the dependent variable in the linear least-squares fit.}
\small
\begin{tabular}{l r r r r}
\tableline
 & HV822 & HV1333 & HV1335 & HV1345 \\
\tableline
$(m-M)_0$ & 18.90 & 19.03 & 19.09 & 18.70 \\
$\sigma(m-M)_0$ &  0.07 &  0.11 &  0.23 &  0.18 \\
      $M_V$ & $ -4.58$ & $ -4.51$ & $ -4.51$ & $ -4.16$ \\
$<M_V>$(PL) & $ -4.68$ & $ -4.65$ & $ -4.50$ & $ -4.42$ \\
$\Delta M_V$ & $  0.10$ & $  0.14$ & $ -0.01$ & $  0.26$ \\
$\Delta M_{V,{\mbox{\scriptsize ridge}}}$(corr) & $  0.30$ & $  0.34$ & $  0.19$ & $  0.46$ \\
\tableline\tableline
\end{tabular}
\normalsize
\end{table}

\section{The effect of metallicity on the P-L relation}

  To determine the effect of metallicity on the zero point of the P-L
relation, we attempt to measure the offset between the P-L relation for
SMC Cepheids and the relation for Galactic Cepheids. As the P-L relation
is a statistical relation and not a direct physical relation, we need to
compare the ridge lines for the two populations. For the Galactic Cepheids
this is already done by Gieren et al. (1998). For the SMC sample we compare
our visual magnitudes with the magnitudes from the sample of Laney and
Stobie (1994), all corrected for depth effects following their method.
We find that our four stars are all on the faint side
of the distribution and we need to add an average offset of $-0.12$~mag
to bring our stars to the SMC ridge line. This together
with the systematic offset described above ($+0.32$~mag) leads to the corrected
magnitude offsets $\Delta M_{V,{\mbox{\scriptsize ridge}}}$(corr) listed
in Table~\ref{tab.FGno_bisec}.

   The weighted average of these offsets is $0.32\pm0.06$ which
we combine with the metallicity offset of $-0.7\pm 0.1$ to obtain $\Delta
\mbox{$M_{V}\,$} / \Delta {\mbox{[Fe/H]}\,} = -0.45\pm0.11$~mag/dex.
To the intrinsic uncertainty we have to add an estimate for the
systematic uncertainties from the corrections performed above. We
estimate this to contribute of the order 0.1~mag giving a final
dependence of $-0.45\pm0.15$. Consequently our preliminary results
shows evidence for a significant metallicity effect in good agreement
with other recent observational results (f.ex. Beaulieu 1997 and
Sasselov et al. 1997).

\acknowledgments We appreciate discussions with Wolfgang Gieren and
Vincenzo Ripepi of earlier versions of the results presented here.

\end{document}